\documentstyle[natbib,emulateapj,onecolfloat]{article}

\newcommand{\Lam}{\Lambda}

\newcommand{\hinv}{h^{-1}}
\newcommand{\mpc}{{\rm\,Mpc}}

\newcommand{\kms}{{\rm\,km\ s^{-1}}}

\newcommand{\Gyr}{{\rm\,Gyr}}

\newcommand{\yrs}{{\rm yrs}}
\newcommand{\yr}{{\rm yr}}
\newcommand{\Msun}{\,M_{\odot}}
\newcommand{\Zsun}{Z_{\odot}}
\newcommand{\himsun}{\hinv{\,\Msun}}
\newcommand{\vv}[1]{{\bf #1}}

\newcommand{\etal}{et~al.}
\newcommand{\ie}{{\frenchspacing i.e.}}
\newcommand{\eg}{{\frenchspacing e.g.}}
\newcommand{\ltsim}{\lesssim}
\newcommand{\gtsim}{\gtrsim}
\def\expec#1{\langle#1\rangle}

\def\bgeqa#1{\begin{eqnarray}\label{#1}}
\def\endeqa{\end{eqnarray}}
\def\Fig#1{Figure~\ref{#1}}

\newcounter{thefigs}

\newcounter{thetabs}

\newcommand{\tableskip}{\tablevspace{3pt}}

\begin{document}

\twocolumn[
\title{Star Formation History and Stellar Metallicity Distribution 
in a $\Lam$ Cold Dark Matter Universe}
\author{Kentaro Nagamine,$^1$ Masataka Fukugita,$^{2,3}$ 
Renyue Cen,$^4$ Jeremiah P. Ostriker,$^4$}
\affil{{}$^1$ Joseph Henry Laboratories, Physics Department, Princeton University, 
Princeton, NJ 08540, USA \\ 
  {}$^2$ Institute for Cosmic Ray Research, University of Tokyo, Kashiwa 2778582, Japan \\
  {}$^3$ Institute for Advanced Study, Princeton, NJ 08540, USA \\
  {}$^4$ Princeton University Observatory, Princeton, NJ 08540, USA
}

\begin{abstract}
We study star formation history and stellar metallicity distribution
in galaxies in a $\Lam$ cold dark matter universe using a hydrodynamic 
cosmological simulation. 
%, and attempt to make predictions which can be used to 
%either falsify or verify the model by future observations. 
%The global star formation rate declines on average with e-folding time of
%$\tau\sim 6\Gyrs$, and intermittent star bursts continue from 
%high-redshift to $z=0$.
%The Madau plot of the simulation rises rapidly from $z=0$ to 2,
%and levels off in the interval $z=4-7$.
%and a moderately increases towards high-redshift thereafter.
%The fractions $(25,50)\%$ of the total stellar 
%mass observed today has formed by $z=(3.6, 1.8)$.
Our model predicts star formation rate declining in time exponentially
from an early epoch to the present with the time-scale of 6 Gyr,
which is consistent with the empirical Madau plot with modest
dust obscuration.
Star formation in $L^*$ galaxies continues intermittently
to the present also with an exponentially declining rate of
a similar time-scale, whereas in small galaxies star formation 
ceases at an early epoch.
The mean age of the extant stars decreases only slowly 
with increasing redshift, and exceeds $1\Gyr$ at $z=3$.
Normal galaxies contain stars with a wide range of metallicity and age:
stars formed at $z<1$ have metallicity of $0.1-1.0\Zsun$, 
while old stars take a wide range of values from $10^{-6}\Zsun$ to
$3.0\Zsun$. The mean metallicity of normal galaxies is in the range 
$0.1-1.0\Zsun$.
Dwarf galaxies that contain only old stars have a wide
range of mean metallicity ($10^{-4}-1.0\Zsun$), but on average
they are metal deficient compared with normal galaxies.
\end{abstract}

\keywords{stars: formation --- galaxies: formation --- cosmology: theory --- methods: numerical}

] %--- for emulateapj.sty -----%

\section{Introduction}
\label{intro_section}
Over the last 15 years, cold dark matter (CDM) models have served as basic frameworks 
to study the formation of cosmic structure, and 
have been successful in delineating a scenario of galaxy formation
\citep{Davis85, Blumenthal84}. We now understand
the formation of large-scale structure reasonably well in terms
of a CDM model dominated by a cosmological constant $\Lam$
%as more and more modern observational data suggest the existence
%of $\Lam$ 
\citep{Efstathiou90, Ostriker95, Turner97, 
%Perlmutter98, Riess98, Bahcall99, 
Balbi00, Lange00, Hu00}.

When and how galaxies formed, however, are a far more complicated 
problem due to  
non-gravitational physical processes operating on small scales. 
Many authors use semianalytic models to study galaxy formation, 
where dark matter halo
formation under the hierarchical structure formation
is supplemented heuristically with physical processes 
for baryons \citep{White91, Baugh98, Kauffmann99a, 
SP99, Cole00}.
They have succeeded to give a picture of galaxy formation 
roughly consistent with observations, 
but have the disadvantage that  one has to assume
a set of simplified model equations, 
often associated with additional free parameters,
for each physical process included in the model.
The advantage of semianalytic models is, on the other hand, 
that they are computationally light, and therefore, 
one can search for a viable model with a small amount of cost.

An alternative but computationally much more expensive 
approach is direct cosmological hydrodynamic simulations.
So far, intergalactic medium and overall galaxy formation 
processes have been well studied with hydrodynamic approach 
\citep[\eg,][]{CO92a, CO93, %CO99a, CO99c, 
CO00, Katz96} \citep[see][however]{Pearce99}.
%probably with a fear that the simulation mesh would not be sufficient to
%study individual galaxies (see Pearce99, however).
Much effort invested in improving 
the accuracy of simulations has brought the hydrodynamic mesh 
approaching to $(1000)^3$, %over the last 10 years,
with which we may hope that 
we obtain meaningful results on some aspects of galaxy properties. 
It is certainly impossible to resolve internal structure of galaxies
with the present simulations. Nevertheless,
we would expect that global properties such as global star formation rate
(SFR) are described reasonably well with a few adjustable parameters, 
since they primarily depend on thermal balance of the bulk of clouds, 
such as how gas cools, 
%in the high density region, 
or how surrounding gas is reheated by feedback by star formation. 
The fact that we obtain a reasonable
amount of baryonic mass frozen into stars and a global metal abundance
with reasonable input parameters
justifies, at least in part, our expectation. 
%, even if the internal structure of galaxies is not resolved.

In this paper, we primarily 
discuss the star formation history of galaxies 
and the evolution of stellar metallicity, which are the direct
output of the simulation. 
The aspects that require additional use of a stellar population 
synthesis model, such as luminosity function and colors of galaxies, 
will be discussed in a separate publication 
\citet[][hereafter Paper II]{Nagamine01}. 
Since the hydrodynamic approach is computationally demanding,
we do not attempt to make a fine tuning of input parameters, but our aim 
here instead is to make qualitative predictions which can be used 
to either verify or falsify the $\Lam$CDM model by future observations
with only a limited number of assumptions, 
treating the dynamics of baryons more accurately than in semianalytic models. 

%Previously, we have discussed the global star formation history 
%using earlier large-scale hydrodynamic simulations \citep{Nagamine00a}. 
%However, the assumed yield of metals was too high in those simulations,
%which resulted in overproduction of stars by about a factor of 2 
%compared to the current empirical estimate by \citet{Fukugita98}. 
%The mass function of galaxies was not reported on
%due to insufficient spatial and mass resolution.
%And while the mass resolution was higher than in other numerical 
%work, it was not high enough for reliable calculations with $z>4$.
%We therefore revisit this important topic with the current
%simulation which has a higher spatial and mass resolution, but
%a smaller volume.  Overall, the present simulation is in good
%agreement with observations as we will describe below. 

Our results would give insight on how galaxies are assembled from
an early epoch to the present. Observationally interesting questions
would include: (i) How old are stars in normal and dwarf galaxies, and what is
their age distribution? A relevant question is whether star formation
takes place continuously or intermittently;  (ii) What is the stellar 
metallicity distribution in normal and dwarf galaxies? A relevant old 
problem is the paucity of metal poor stars in the solar neighborhood 
\citep[G-dwarf problem:][]{Bergh62, Schmidt63}; (iii) Is there
a unique age-metallicity relation for stars in normal galaxies, and 
(iv) Is there a relation between the mass of galaxies and average 
metallicity? How these aspects evolve with redshift is also an
observationally relevant problem, although the observational studies 
could answer only for global properties for high-redshift objects.
Our simulation yield unambiguous predictions to these problems, 
at least qualitatively.

%The latter aspect concerns the age-metallicity 
%relation of stars. Most chemical evolution models (e.g., Lynden-Bell 
%(1975); Pagel 1989) predicts a
%well-defined age-mass relation of galaxies. We study the
%consequence that will be resulted from the CDM galaxy formation on this
%problem. The related issue is the G-dwarf problem of the Galactic disk
%(van den Bergh 1962; Pagel \& Pritchett 1975). While
%we cannot resolve disk and bulges in our simulation, the general
%picture obtained from simulations shed light how these problems
%would be solved in the CDM model. 

In \S~\ref{simulation_section}, we describe our simulation.
In \S~\ref{sf_section} we first present the global star formation history 
of the entire simulation box. We then discuss in \S~\ref{divide_sf_section}
the star formation history of individual galaxies.
The mean age of the stars and the formation epoch of galaxies 
are discussed in \S~\ref{age_section}.
We discuss the stellar metallicity distribution and 
the mean metallicity of galaxies in \S~\ref{metal_section},
and conclude in \S~\ref{conclusion}.

\section{Simulation and Parameters}
\label{simulation_section}
We use a recent Eulerian hydrodynamic cosmological simulation with a
comoving box size of 25$h^{-1}$ Mpc with $768^3$ grid cells and 
$384^3$ dark matter particles of mass $2.03\times10^7\himsun$. 
The comoving cell size is $32.6h^{-1}$ kpc, 
and the mean baryonic mass per cell is 
$3.35\times10^5h^{-1}M_\odot$ ($h$ is the Hubble constant in units
of $H_0=100\kms\mpc^{-1}$).
The cosmological parameters are chosen to be 
$\Omega_m=0.3$, $\Omega_{\Lam}=0.7$, $\Omega_b h^2=0.016$, $h=0.67$, 
$\sigma_8=0.9$, and  the spectral index of the primordial 
mass power spectrum $n=1.0$. 
The present age of the universe is 14.1 \Gyr. 
The structure of the code is similar to that of 
\citet{CO92a, CO93}, but significantly improved over the years.

The Eulerian scheme we use here 
has higher mass resolution than other hydrodynamic approaches, such as
the smoothed particle hydrodynamics \citep[SPH; \eg,][]{Katz96, Dave97}
and the adaptive mesh refinement \citep[AMR; \eg,][]{Bryan95, Kravtsov97},
while SPH and AMR usually have higher spatial resolution 
than the Eulerian method.
We remark that our code is specially designed to capture the 
shock region well, using the total variation diminishing method
\citep{Ryu93}.
The higher mass resolution means more resolving power
at higher wavenumbers for a given power spectrum, and hence
a better treatment of early structure formation.

The baryons are turned into stars in a cell of 
overdensity $\delta\rho/\rho>5.5$
once the following three conditions are satisfied: (i) contracting flow
$\nabla\cdot\vv{v}<0$, (ii) fast cooling $t_{\rm cool}<t_{\rm dyn}$, and 
(iii) Jeans instability $m_{\rm gas}>m_{\rm J}$.  
Each stellar particle has a number of attributes at birth, including 
position, velocity, formation time, mass, and metallicity.
Upon its formation, the mass of a stellar particle is determined by 
$m_{\ast}=c_{\ast} m_{\rm gas} \Delta t/t_{\ast}$, 
where $\Delta t$ is the current time-step in the simulation, and 
$m_{\rm gas}$ is the baryonic gas mass in the cell. We take 
$t_{\ast}={\rm max}(t_{\rm dyn}, 10^7\yrs)$, where 
$10^7 \yrs$ is the shortest time-scale of star formation for 
O stars.
%, while we cannot resolve dynamics shorter than the dynamical time-scale. 
%, as seen in some Local Group dwarf galaxies. 
We assume the star formation efficiency to be $c_{\ast}=0.25$.
The fraction of baryonic mass that collapses into dense
state is fairly well defined by our code, nearly independent of 
the detailed values of $c_{\ast}$ and the minimum value of $t_{\ast}$,
as confirmed by a recent numerical experiment of Norman
(2000, private communication).
%The following analogy helps to understand
%this. When one drains the water from a bath tub, the amount of time
%it takes depends on the size of the hole ($c_{\ast}$ and $t_{dyn}$) 
%of the tub, but the total amount of water (total cooled gas) 
%that you drain does not change according to the size of the hole. 

The mass of the stellar particles ranges from 
$10^{3.5}-10^{10.3}\himsun$; with the mean 
$\expec{m_*}=10^{6.9}\himsun$ at $z=0$.
An aggregation of these particles is regarded as a galaxy.
%(for grouping, see below). 
The stellar particle is placed at the center of the cell after its 
formation with a velocity equal to the mean velocity of the gas,
and followed by the particle-mesh code thereafter as collisionless 
particles gravitationally coupled with dark matter and gas.

Feedback processes such as ultra-violet (UV) ionizing field, 
supernova (SN) energy, 
and metal ejection are also included self-consistently.
The SN and the UV feedback from young stars are treated as follows: 
$\Delta E_{\rm SN} = m_{\ast} c^2 \epsilon_{\rm SN}$
and $\Delta E_{\rm UV}=m_{\ast} c^2 \epsilon_{\rm UV} g_{\nu}$, 
where $g_{\nu}$ is
the normalized spectrum of a young, Orion-like stellar association 
taken from \citet{Scalo86}, and the efficiency parameters are taken as
$(\epsilon_{\rm SN},\epsilon_{\rm UV})=(10^{-5}, 3\times10^{-6})$ 
\citep[cf.][]{CO93}.
%, where the value of $\epsilon_{\rm SN}=10^{-5}$ 
%corresponds to the ejecta velocity of $\approx 2700 \kms$ for the ejecta
%mass fraction given below. 
The $\Delta E_{\rm SN}$ is added locally in a cell, and 
the $\Delta E_{\rm UV}$ is averaged over the box.
Extra UV is coadded for quasars as
$\Delta E_{\rm AGN}=m_{\ast} c^2 \epsilon_{\rm AGN} f_{\nu}$, where 
$f_{\nu}$ is an AGN spectrum taken from \citet{Edelson86},
and $\epsilon_{\rm AGN}=5\times10^{-6}$ is adopted.
The proportionality between $\Delta E_{\rm AGN}$ and $m_{\ast}$ 
is an assumption, and our ignorance is included in 
$\epsilon_{\rm AGN}$ which is adjusted to fit the 
hard X-ray background observed by ASCA 
\citep[see][]{Phillips00}.
Metals are created according to $m_Z = Y m_{\ast}$, 
where $m_Z$ is the mass of metals, and $Y$ is the yield. 
The fraction 25\% ($f_{ej}=0.25$) of the initially collapsed baryons 
is ejected back into the intergalactic medium in the local grid cell.
This ejected gas is polluted by metals, and 
2\% \citep[$Y=0.02$;][]{Arnett96} of initially collapsed baryons
are returned to intergalactic medium as metals, which are followed 
as a separate variable by the same hydro-code which follows the gas density. 
 The value of the three parameters ($\epsilon_{\rm SN}$, $\epsilon_{UV}$, $Y$) 
are proportional to the amount of high-mass stars for a given amount of 
collapsed baryons. 
The results presented in this paper is insensitive to the assumed
value of $f_{ej}$ provided that it is non zero.
Thus, all our ignorance concerning star formation is parameterized 
into essentially one free parameter, the efficiency of cooling, i.e., 
collapsing matter that is transformed into high-mass stars,
and this value is empirically calibrated so that one obtains 
a final cluster gas-metallicity of $Z=\frac{1}{3}\Zsun$, 
where $\Zsun=0.02$ is the solar metallicity.

We use the HOP grouping algorithm \citep{Eisenstein98} to identify galaxies
with threshold parameters $(\delta_{\rm outer},\delta_{\rm saddle},
\delta_{\rm peak})=(80,200,240)$.
We set the minimum number of grouping particles to 5; 
changing this to 2 does not modify the catalog.
We identify 2097 galaxies in the entire box at $z=0$, 
the minimum galaxy stellar mass being $10^{6.7}h^{-1}\Msun$.
The total mass of stars corresponds to $\Omega_{\ast}=0.0052$, \ie, 
15\% of baryons are collapsed into stars 
(including the ejected gas). 
This is consistent with the upper limit of an empirical estimate of 
\citet{Fukugita98}.
The average metal density is $10^{7.1}\Msun\mpc^{-3}$, 
which is also close to the empirical estimate of $10^{7.4}\Msun\mpc^{-3}$
(Fukugita \& Peebles, unpublished), justifying our choice of parameters.

We examine the merger history of galaxies, 
and find that 93\% of the galaxies at $z=0.5$ preserve more than 80\% of 
their stellar mass to $z=0$: \ie, most of the constituent 
stellar particles are not lost upon merging or tidal shredding. 
This tells us that the galaxies identified in our simulation 
are dynamically stable enough to obtain 
meaningful results for the evolution of individual galaxies. 
%A visual impression of how the grouped stellar particles, \ie ``galaxies'',
%are distributed with in dark matter halos can be seen in 
%Figure 1 of \citet{CO00} from our earlier simulation.

\section{Global Star Formation Rate}
\label{sf_section}
We first present in Figure~1a the global SFR as a function 
of time, going from right to left. The prediction of global SFR
in the CDM model has been discussed in the literature
\citep*{Baugh98, Nagamine00a, SP01} 
\footnote{The simulation used by \citet{Nagamine00a} assumed 
a too high yield of metals ($Y=0.06$), which resulted in 
overproduction of stars compared to the empirical estimate. 
Also, the simulation mesh was coarser and one may suspect 
the effect of low resolution at above $z\approx 4$. 
We consider that the result presented in this paper
supersedes that of \citet{Nagamine00a}}.
%
% the previous $\lbox=50\himpc$ simulation 
%had lower mass and spatial resolution than the current one, 
%as well as a different set of cosmological parameters 
%($\Omega_m=0.37$, $\Omega_{\Lam}=0.63$) with $\Omega_b=0.049$ and $Y=0.06$, 
%and resulted in $\Omega_{\ast}=0.018$, which 
%is larger than the upper limit of the empirical estimate of
%\citet{Fukugita98} by about a factor of 2. 
%Therefore, we consider the present simulation to be 
%more reliable and closer to the reality than the previous simulations, 
This quantity is observationally known up to uncertainties 
associated with dust obscuration, and the agreement with 
observations can be used as a verification of the calculation.
%, especially of clustering and gross thermal balance. 
This quantity does not receive the effect of the grouping procedure, 
but gives a measure of the efficiency of gas cooling, 
which depends on the thermal balance.

We also indicate a boxcar smoothed histogram with 
10 point running average with the dotted curve. 
The dashed line is a least-square fit 
to the exponential decay function, 
corresponding to a time-scale of $\tau=5.9\Gyr$.

%The earliest peak of the star formation occurs before $z=5$, 
%and spiky starbursts continue up to the present epoch 
%as galaxies accrete gas and merge into larger systems.
%The box size of $\lbox=25\himpc$ is not large enough to 
%obtain a smooth cosmic star formation history, but our result indicates
%that the star formation history of a volume of $(25\himpc)^3$ 
%in our universe could be as violent as we see 
%in this simulation with successive starbursts. 

\begin{figure}[htb]
\centerline{\epsfxsize=3.4truein\epsffile{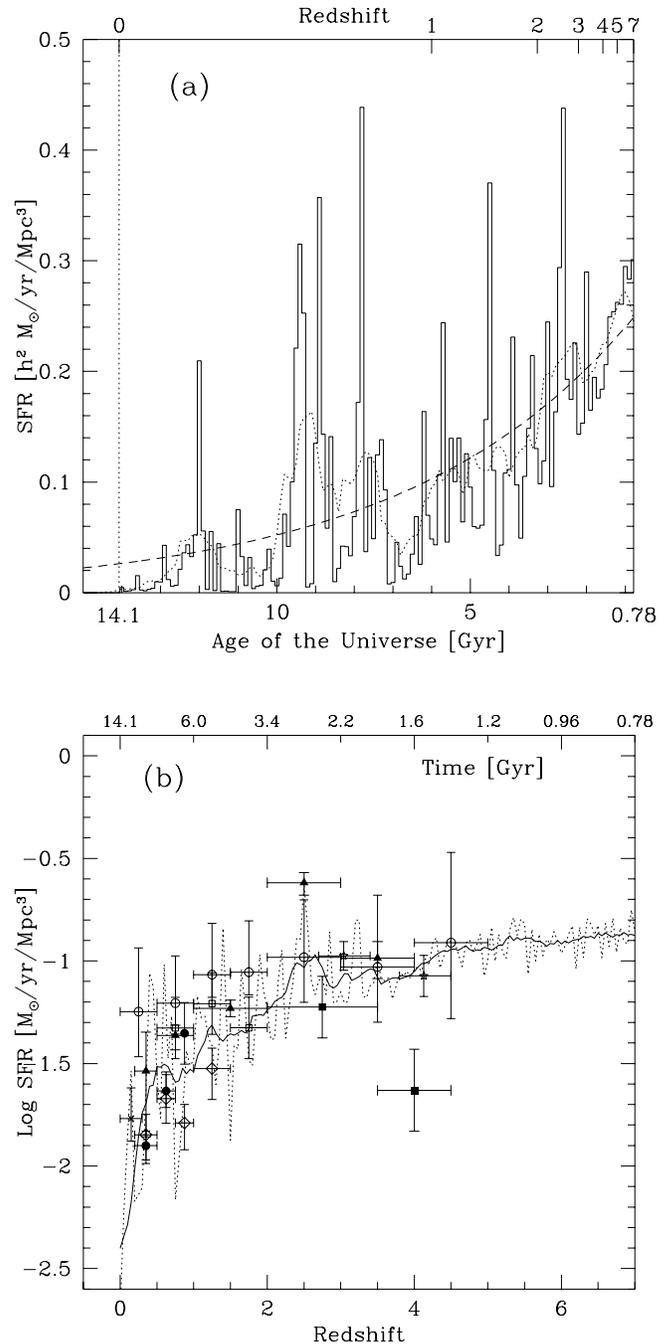}}
\caption{\footnotesize
(a) The SFR of the entire simulation box
as a function of time, going from right to left.
The dashed line is the least-square fit to the decaying exponential 
function given in Table~\ref{table1} with a time-scale of $\tau=5.9$\Gyr. 
(b) The identical information, the SFR, as a function of redshift (the Madau plot).
The source of data points are given in \citet{Nagamine00a}. 
Here we adopted factors of 1.3 ($z<2$) and 2.5 ($z>2$) 
for dust extinction correction of the observed data points. 
See text for discussion.}
\label{total_SF.ps}
\end{figure}

\begin{table}[tb]
\footnotesize
\caption{\label{table1}}
\begin{center}
{\sc Parameters of Exponential Fit for SFR}
\begin{tabular}{ccc}
\tableskip\hline\tableskip
Mass range [$h^{-1}\Msun$] & A [$h^2\Msun\yr^{-1}Mpc^{-3}$] & $\tau$ [Gyr] \\
\tableskip\hline
All & 0.24 & 5.9 \\
$M_{\rm stellar}>2\times10^9$ & 0.23 & 6.1 \\
$2\times10^8<M_{\rm stellar}<2\times10^9$ &  0.021 & 0.83 \\
$M_{\rm stellar}<2\times10^8$ & 0.011 & 0.55 \\
\tableskip\hline
\end{tabular}
\end{center}
NOTES.---%
{Parameters given above are the least-square-fit of the SFR 
in Figure~\ref{total_SF.ps} and \ref{divide_SF.ps} to the following 
exponential function: 
${\rm SFR}~[h^2\Msun \yr^{-1} Mpc^{-3}] = A~exp(-t/\tau)$, 
where A is the SFR at $t=1\Gyr$, and $\tau$ gives the e-folding time
of star formation. The rise in the first 1~Gyr is neglected upon fitting. 
}
\end{table}

The identical information is presented in
Figure~1b, but using the logarithmic scale for SFR and redshift 
as a time-scale in accordance with the Madau plot. In this diagram
the meaning of solid and dotted curves are reversed:
the dotted curve follow the raw data, and the solid line is
the smoothed data. The observation is corrected for 
dust extinction 
%(also corrected to our $\Lam$CDM cosmology)
according to the prescription of \citet{Steidel99}: 
we assume highly uncertain extinction correction factors to be 1.3 ($z<2$) 
and 2.5 ($z>2$) to obtain a better fit, while \citet{Steidel99} 
used higher values 2.7 ($z<2$) and 4.7 ($z>2$). 
With a rather large error in the empirical estimate of SFR,
a global agreement is seen between our calculation and the observation.
We do not observe a peak of SFR at low redshift, as often found by
semianalytic modellers of galaxy formation 
\citep{Baugh98, SP01}
\footnote{In semianalytic models, 
one may suppress star formation at high-redshift by taking 
a larger SN energy feedback parameters. 
In general, they require a strong SN feedback to fit the 
faint-end of the galaxy luminosity funcition \citep[see][]{Cole00}.
See Paper II for the luminosity function in our simulation.}.
The SFR is on average a smooth function of redshift, 
and nearly levels off at $z>4$.  
Note that this behavior is consistent with an exponential
decay of the SFR in time.
This means that the bulk of stars have formed at redshift higher 
than 2: 25\% of stars formed at $z>3.6$, and another 25\% were added
between $z=3.6$ and 1.8 (By $z=1$, 68\% of stars were formed). 
We consider that
our result is reliable up to $z=5-6$ from a mesh effect consideration, 
whereas that at a yet higher redshift
may receive an effect of poor resolutions.
Observationally it is an unsettled problem whether SFR levels
off at high-redshift \citep{Steidel99} or still increases beyond
$z>3$ \citep{Lanzetta99}. 
It is an interesting observational problem to find out how SFR behaves
at high-redshifts, but this can be done only after proper knowledge of
dust extinction or with the observations that do not receive effects
of extinction. For now, it is sufficient for us to know that the calculation
of global SFR is grossly consistent with the observation.

\section{Star Formation History in Galaxies}
\label{divide_sf_section}

\begin{figure}[htb]
\centerline{\epsfxsize=3.4truein\epsffile{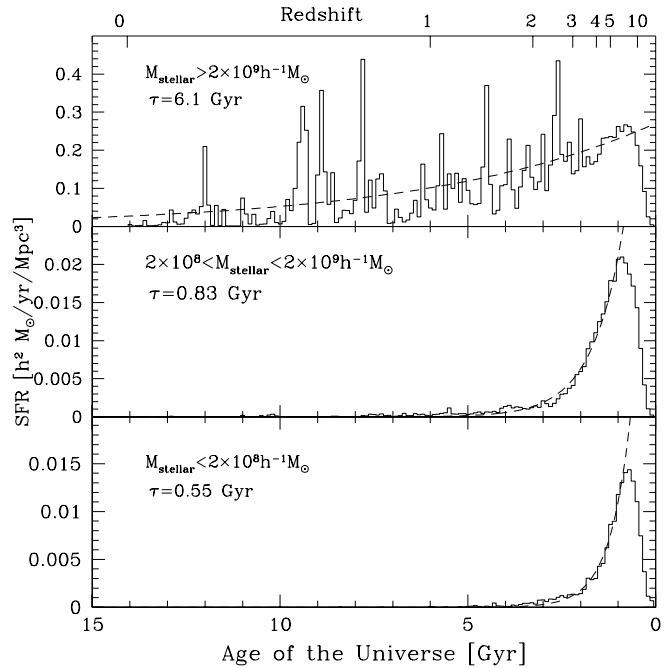}}
\caption{\footnotesize
Star formation history of the three different samples of galaxies
divided by their stellar mass at $z=0$. 
Note the different scales on the ordinate.
The dashed lines are the least-square fits to the exponential 
function given in Table~\ref{table1}.
Massive galaxies (top panel) continue to form stars until 
the present epoch, while less massive galaxies (bottom two panels) stop 
forming stars at higher redshift. 
}
\label{divide_SF.ps}
\end{figure}

In Figure~\ref{divide_SF.ps}, we present the star formation histories 
of galaxies in the simulation divided into three samples according to 
their stellar mass at $z=0$.
Note that the starbursts at high-redshift may have taken place in 
smaller progenitors that later merged into a 
massive object that is classified in the largest mass bin.
All galaxies which fall into each mass interval are 
co-added in the histogram. 
Most of the mass reside in large galaxies which are represented 
in the top panel.
This histogram can also be interpreted as the age
distribution of stars in galaxies. 
The dashed lines show the least-square fits to 
the exponential decay function with parameters given in 
Table~\ref{table1}. %with the time-scale of 
%$\tau=6.1, 0.83$, and 0.55$\Gyr$,  from top to 
%bottom, respectively.
The time-scale $\tau=6.1$ Gyr for `normal galaxies' (stellar mass 
$M_{\rm stellar}>2\times10^9\himsun$; top panel of Figure~2)
agrees qualitatively with the star formation history inferred 
for present-day Sbc galaxies, 
which are about the median type of local galaxies 
\citep{Sandage86, Gallagher84}.
We show in Paper II that the time-scale $\tau=6.1\Gyr$
gives the correct mean color of normal galaxies.
For these galaxies, 
36\% of their total stellar mass formed below $z=1$, 
23\% from $z=1$ to 2, and 40\% at $z>2$.

A feature dominant in the top panel is the strong spikes 
in the star formation history, indicating that the star formation
takes place intermittently rather than continuously in normal galaxies. 
This feature is more manifest if we look at the star formation 
history of individual galaxies. 
Each galaxy experiences several conspicuous star
formation episodes, while they are quiescent for most of the time.
This agrees with the suggestive evidence from the study of stellar ages
in the Milky Way that it has experienced enhanced episodes of
star formation in the past \citep{Majewsky93, Rocha-Pinto00b}.
A study of star forming galaxies also supports the view that 
star formation is episodic \citep{Alonso-Herrero96}.

The figure (lower panels) also shows that star formation
ceased at high-redshift in small galaxies.
Only 7\% of stars form below $z=1$ in galaxies with 
$2\times 10^8<M_{\rm stellar}<2\times10^9h^{-1}\Msun$, and
no stars form below $z=1$ in galaxies with  $M_{\rm stellar}<2\times 10^8$.
This is consistent with the feature generic to the CDM structure 
formation, in which the formation of smaller structure ceases
at higher redshift. Accretion of material stops, while most of
residual gas is easily swept out by supernovae and photoionization
from small systems
\citep{Dekel86, Rees86, Efstathiou92, Quinn96, Navarro97, Weinberg97, 
Gnedin00b}.
Our prediction is that, in a CDM universe, small galaxies found 
today are dominated by old stars; 
dwarf galaxies consist dominantly of stars of 10~Gyr 
old and do not show star formation activity in the recent epoch. 
%We consider that 
%this is a feature generic to the CDM model: small objects
%formed only at high redshift.  
It is known that most of dwarf galaxies in the Local Group  
are dominated by old stellar population with occasional mixture
of younger stars \citep{Grebel97, Mateo98, Bergh98, Gnedin00a}. 
A small admixture of young stars may not be excluded in the
CDM model, since our calculation adopts a simplified treatment
where only the evolution of the bulk of the material is followed,
and it has no power to resolve minor components. 
The bulk star formation in small systems at late times, 
however, cannot be understood within the the scenario of 
hierarchical galaxy formation based on the CDM model.
Major star formation activities in recent epochs,
such as those in Leo I and Carina, do not fit the model
\footnote{We note in passing that, in the warm dark matter scenario, 
low mass galaxies form primarily by ``top-down'' fragmentation 
rather than ``bottom-up'' accumulation process and so are found 
later than in the standard CDM picture \citep{Bode00}.}.

%Another important and interesting question we should ask is, 
%what stops the star formation in those less massive galaxies 
%after $t=3\Gyr$? There are several possible reasons for this:
%(1) They have exhausted their gas fuel in the first burst of star
%formation which took place at $t<3\Gyr$.
%(2) The gas was blown out of small dark matter halos by 
%SN feedback in the first episode of star formation.
%(3) Further star formation is suppressed by the photoionizing
%UV background.
%(4) It is due to selection: low-mass systems in high-density
%regions have merged out of existence so those that remain
%live in low-density regions (``voids'') where mergers and accretion
%become rare.
%(5) The simulation does not have enough spatial resolution 
%at late times to resolve further infall of gas into these small systems.
%To find out the answer to this question, 
%more detailed investigations of gas in these systems 
%are required, as have been done by several authors 
%\citep{Quinn96, Navarro97, Weinberg97, Gnedin00b}.
%Explanation (4) could be tested by looking at 
%the decline rate $\tau$ of star formation as a function local density.
%However, these questions are beyond the scope of this paper, 
%and we plan to investigate these issues more in the future.

\section{Mean Age of Stars and Galaxies}
\label{age_section}
\begin{figure}[htb]
\centerline{\epsfxsize=3.4truein\epsffile{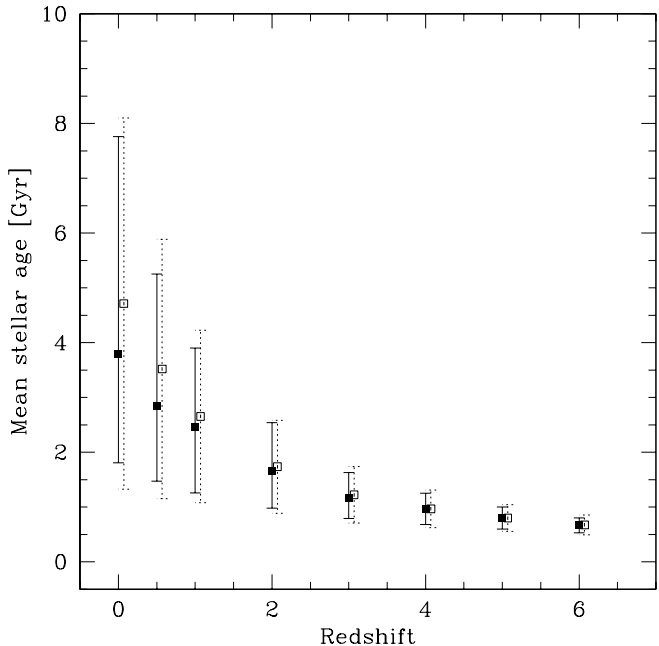}}
\caption{\footnotesize 
Median (solid squares) and the mass-weighted mean stellar age 
(open squares) of all the stars at each epoch. The solid bars show 
the quartiles, and the dotted bars show 1-$\sigma$. 
The median and the quartiles were obtained from the cumulative 
distribution of the stellar mass as a function of the age at each epoch.
The mean stellar age decreases slowly with increasing redshift, 
and exceeds 1~Gyr at $z\approx 3$.}
\label{star_age.ps}
\end{figure}

The mean stellar age at each epoch in the history of the universe is also 
interesting. In Figure~\ref{star_age.ps}, we plot the median (solid square) 
and the mass-weighted mean age (open square) of all stars at each epoch.
The median was calculated from the cumulative distribution of 
stellar mass at each epoch.
The solid bars indicate the quartiles, which almost coincide with 
1-$\sigma$ ranges shown with dotted bars.
The average age of the extant stars decreases only slowly with 
increasing redshift. The mean stellar age today is $4-5 \Gyr$ and the age
becomes smaller than 1~Gyr only at $z=3$. 
The age $4-5\Gyr$ is slightly younger than the observed 
median stellar age of $\approx 6\Gyr$ in the Milky Way \citep{Rocha-Pinto00b}.
We show in Paper II that the distribution of $B-V$ colors of 
simulated galaxies agrees with that of the local galaxy sample. 
The median age of 1~Gyr at $z=3$ means 
that the majority of galaxies at this redshift are 
dominated, in addition to star forming activities, 
by Balmer absorption features of A stars, which would 
be tested with near infrared spectroscopy for Lyman break galaxies.

\begin{figure}[b]
\centerline{\epsfxsize=3.4truein\epsffile{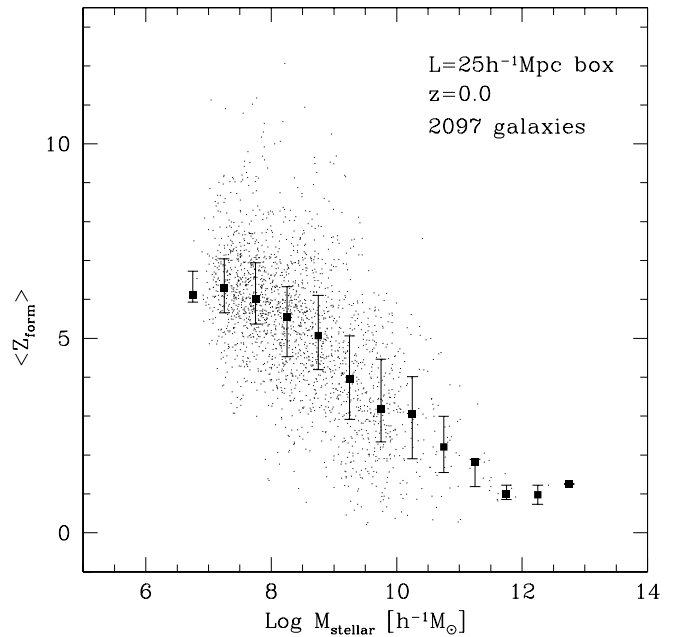}}
\caption{\footnotesize 
Mean formation time (converted to redshift) of galaxies 
vs. stellar mass of galaxies at $z=0$ in the simulation.
Mean formation time is calculated by taking the mass-weighted 
average of formation time of constituent particles of each galaxy.  
The solid square symbols show the median and the solid bars show 
the quartiles in each mass bin. 
Less massive galaxies formed earlier, and later merged into massive ones.
}
\label{age_mass.ps}
\end{figure}

Figure~\ref{age_mass.ps} gives the mean formation 
redshift of galaxies  versus their 
stellar mass at $z=0$. The mean formation redshift was calculated 
by taking the mass-weighted mean of the formation epoch (in time unit) 
of constituent stellar particles, and then converting it to redshift.
Each dot represents individual galaxy.
The solid square symbols are the median, and the solid 
bars are the quartiles in each mass bin. 
This figure confirms the result we saw earlier that
less massive galaxies found at $z=0$ are older 
than massive ones. 
Small systems formed at high and intermediate
redshift merge into more massive systems at lower redshift.
The mean formation redshift mostly decreases
with increasing galaxy stellar mass, but it
turns over slightly at the very massive end:
this is due to the merging of oldest dwarfs into very massive galaxies.

\section{The metallicity distribution and the age-mettalicity relation}
\label{metal_section}
\begin{figure}[t]
\centerline{\epsfxsize=3.4truein\epsffile{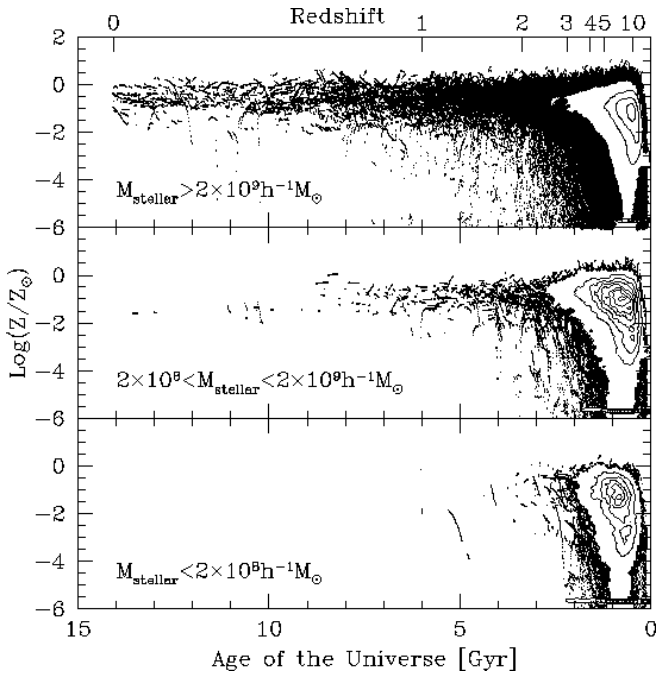}}
\caption{\footnotesize 
Metallicity distribution of stars for the same subsamples as in
Figure~\ref{divide_SF.ps}. Each point in the figure 
corresponds to one stellar particle in the simulation with different 
masses.
The abscissa is the age of the universe going from right to left, 
and the contours are of the number density distribution of stellar 
particles on an equally spaced logarithmic scale. 
The mass-weighted mean metallicity at each redshift is given 
in \S~\ref{metal_section}. 
Stars formed earlier ($z>4$) have wide range of metallicity ranging 
from $10^{-6}-1.0\Zsun$, while those formed at $z<1$ have roughly 
$0.1-1.0\Zsun$.
}
\label{divide_metal.ps}
\end{figure}

The metallicity of stars as a function of their 
formation time (going from right to left) is plotted in 
Figure~\ref{divide_metal.ps} for the same subsamples 
as in Figure~\ref{divide_SF.ps}. 
Each point in the figure corresponds to one stellar particle 
in the simulation.
The contours of an equally spaced logarithmic scale in number density
is used where the density of data points is large.
The plot shows that the stars which formed at 
low-redshift ($z\ltsim 1$) do not show unique metallicity but 
distribute from $0.1\Zsun$ to $1.0\Zsun$.
The width of the distribution slowly increases with redshift, 
while the upper envelope stays at 1.0$\Zsun$ up to $z\approx 1$. 
The widening trend of the age dispersion is conspicuous at 
redshift $z>2$. Note that the widening takes place dominantly 
towards lower metallicity, while the upper envelope slightly
increases to super-solar values.
%while the upper envelope is nearly constant. 
At $z\gtsim 3$, the distribution ranges
from $10^{-6}\Zsun$ to $3.0\Zsun$.
It is interesting to observe that the median of metallicity changes 
very little at $z\ltsim 2$. The mass-weighted mean metallicity
varies also slowly as $0.55, 0.45, 0.61, 0.24$, and 0.22$\Zsun$ 
for $z=0, 1, 2, 3$, and 5, respectively. The decrease of mean metallicity
above $z=2$ is clear, whereas we can not conclude the change 
of mean metallicity for $z\leq 2$ because of small statistics.

It is known that stars in the Milky Way do not obey a unique
age-metallicity relation, but show a significantly wide distribution
ranging from $0.1\Zsun$ to $3\Zsun$.
Whether the dispersion increases with redshift, however, 
is still a matter of debate. 
\citet[][see also \citet{McWilliam97}]{Edvardsson93} argued
that the dispersion increases as stellar age gets older, while the
metal rich stars form an envelope that is about solar metallicity 
independent of stellar age. On the other hand, \citet{Rocha-Pinto00a}
argued that age and metallicity show a stronger correlation,
earlier stars being metal poor. Our result agrees with 
Edvardsson's, but not with Rocha-Pinto's result.

The result we discussed above means that early galaxies are not 
necessarily metal deficient.  \citet{Kobulnicky99} show that
emission line galaxies at $z=0.1-0.4$ have metallicity only slightly
lower than solar, which is not uncommon in local galaxy samples. 
\citet{Pettini00} show that Lyman break galaxies at $z\approx 3$
have $\approx 0.25\Zsun$, only modestly sub-solar. 
The strength of metal lines in the
highest redshift quasar \citep{Fan00} is also suggestive of normal
metallicity at $z=5.8$. These observations are all consistent with our
simulation.

It is also our prediction that the Milky Way contains highly metal poor
stars. We expect that 1\% of stars have $\ltsim 10^{-4}\Zsun$. 
Those stars are necessarily old. Another interesting
result is that most super-metal rich stars are from high redshift epoch. 
This would be consistent with the fact that most of super-metal rich stars 
are in the bulge rather than in the disk \citep{Frogel87, Rich88}
if the bulge contains more old stars than the disk
as usually conceived\footnote{A contradictory view is presented by
\citet{Rocha-Pinto00a}, who claim that super-metal rich stars are young}. 
At the same time, our calculation predicts 
that the mean metallicity of the old population is lower
than that of the younger population, which is also consistent with
the observation \citep{Edvardsson93, McWilliam94, Rocha-Pinto00a}.

We may ascribe the change of scatter in metallicity from  
high to low redshift to the following reason. Metallicity is a strong 
function of local overdensity, as shown 
by \citet{Gnedin98} and \citet{CO99c}. 
Both authors have shown that the metallicity distribution 
in high-density regions is narrower than that in low-density
regions. 
%, with the value of $0.1 - 1.0 \Zsun$. 
As time progresses, the universe becomes more clumpy,
and star formation takes place only in moderate to 
high overdensity regions where gas is already polluted by 
metals, resulting in a narrower scatter
of metallicity at late times as seen in the top panel 
of \Fig{divide_metal.ps}.

%It is interesting to note that many early stars formed 
%at $z\sim 5$ have 0.1$\Zsun$ or even larger. 

Let us now consider the G-dwarf problem; the fact that there are 
too few metal poor stars in the solar neighborhood compared 
to the closed box model \citep{Bergh62, Schmidt63}.
In our model, a large fraction of stars formed in an early epoch, 
and such a population contains a significant fraction
of metal poor stars. 
A histogram analysis of the metallicity distribution at
$z=0$ shows that 50\% of stars have $Z<0.3\Zsun$ in our simulation.  
This is roughly the same as the prediction of the closed box model, 
whereas the observation shows the fraction of these stars to be 
only a few percent \citep{Pagel75, Sommer-Larsen91}. 
The qualitative trend of our prediction does not seem to be 
easily modified. We ascribe, however, the presence
of large metal poor stars in our simulation to our inability of 
resolving the internal structure of galaxies (\ie disk from bulge). 
If the disk is a later addition to the bulge 
as the observations suggest \citep{Fukugita96},
it would contain little metal poor stars,
as is clear from \Fig{divide_metal.ps} above. 
The material accreted onto galaxies at low-redshift was
already polluted by metals from bulge stars \citep{Ostriker75}.
%This picture is in agreement with the observation that 
%the disk is a young entity, as inferred from $B-V$ color and the HI content.
A high fraction of metal poor stars could occur in the CDM model, 
when we do not distinguish the disk and the bulge components.  
This view is consistent with the work by \citet{Kauffmann96}, 
where she provides a natural explanation to the G-dwarf 
problem, using a semianalytic model, by pre-enrichment
by early outflows from dwarf progenitors.

%For the top two panels, $(50,95)\%$ of the stars have 
%$Z<(0.3, 1.5)\Zsun$.
%For the intermediate mass range (2nd bottom panel), 
%$(50,95)\%$ of the sample have $Z<(0.08, 0.5)\Zsun$. 
%And for the least massive galaxies, $(50,95)\%$ stars 
%have $Z<(0.03, 0.3)\Zsun$.
%For the top panel, the stellar mass fraction that has metallicity
%less than $(10^{-5}, 10^{-4}, 10^{-3}, 10^{-2}, 10^{-1}, 1.0)\Zsun$ is 
%$(0.7, 1.1, 1.9, 4.0, 20.0, 88.5)\%$.
%In the 2nd top panel, we also show the case for
%$M_{\rm stellar}>2\times 10^{11}\hinv\Msun$, in which
%case 33\% of the stellar mass in those galaxies have
%$Z<0.25\Zsun$ (the fraction is 43\% for the total stellar 
%mass distribution). 

%\begin{figure}[t]
%\centerline{\epsfxsize=3.4truein\epsffile{f7.eps}}
%\caption{\footnotesize 
%Cumulative stellar mass fraction which has metallicity smaller 
%than the given metallicity. The top panel is for all stars,
%and the bottom three panels are for the same subsamples as in 
%Figure~\ref{divide_SF.ps}.
%The dotted lines indicate the metallicity below which 50 and 95\% 
%of the stars is contained, whose exact values are given in 
%\S~\ref{metal_section}. 
%Stars in less massive galaxies tend to have lower metallicity, 
%and take more broadly distributed values as shown in 
%Figure~\ref{mass_metal.ps} more explicitly. 
%The dashed line in the 2nd top panel is for the case of
%$M_{\rm stellar}>2\times 10^{11}\hinv\Msun$. 
%See text for the discussion on the G-dwarf problem. }
%\label{cum_metal.ps}
%\end{figure}

\begin{figure}[htb]
\centerline{\epsfxsize=3.4truein\epsffile{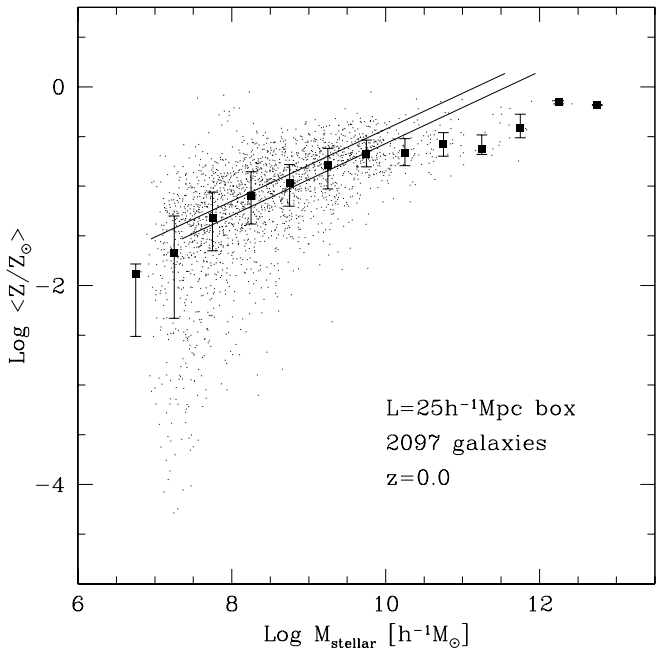}}
\caption{\footnotesize 
Average metallicity of galaxies vs. mean stellar mass
at $z=0$ in the simulation. Each point represents one galaxy. 
The median in each mass bin is shown as a solid square, 
and the solid bars with it are the quartiles.
The two solid lines in the figure are obtained from 
the best-fit to the observational data of spiral and irregular galaxies 
given in Figure 4 of \citet{Kobulnicky99}. See text for details. 
Massive galaxies ($M_{\rm stellar}>10^{10}\himsun$) have normal metallicity 
of $0.1-1.0\Zsun$, while less massive galaxies take a wide range 
of values ($10^{-4}-1.0\Zsun$). 
}
\label{mass_metal.ps}
\end{figure}

Finally in Figure~\ref{mass_metal.ps}, 
we show the mean metallicity of galaxies 
as a function of their stellar mass in the simulation. 
The mean metallicity of each galaxy is calculated by taking the 
mass-weighted average of constituent stellar particles.
Solid square points are the median, and the solid bars are 
the quartiles in each mass bin.
Massive galaxies ($M_{\rm stellar}>10^{10}\himsun$) at $z=0$ 
have mean metallicity of $0.1-1.0\Zsun$, while 
less massive galaxies have a wide range of 
mean metallicity ranging from $10^{-4}-1.0\Zsun$.
There are a few dwarf galaxies which
have even $10^{-6}\Zsun$ 
outside of the figure. 
The star formation in less massive galaxies 
(identified at $z=0$) took place with a wide range of 
metallicity at high-redshift as we saw in the bottom 
panel of Figure~\ref{divide_metal.ps}, 
while larger galaxies at $z=0$ had star formation 
continuing up to the present epoch with 
normal metallicity of $0.01-1.0\Zsun$. 
The two solid lines in the figure are obtained from 
a fit to the observational data of spiral and irregular galaxies 
given in Figure 4 of \citet{Kobulnicky99}.
We converted the absolute $B$-magnitude to the stellar mass 
using the relation $\log M_{\rm stellar}=-0.4 M_B + (2.75\pm 0.20)$,
which the majority of galaxies in our simulation follow 
before dust extinction (see Paper II), and
%(luminosity is obtained using a population synthesis code,
%which results will be reported on in separate papers including Paper II).
scaled the metallicity by $\log (Z/\Zsun)=12+\log (O/H)-8.89$. 
A good agreement is seen between the simulated result and 
the observation in the range 
of $10^7<M_{\rm stellar}<10^{10}\hinv\Msun$.
We remark that \citet{Kauffmann96} obtains a result similar
to ours, which again suggests that 
the reason we are not seeing a similar cumulative 
distribution to her result and the solar neighborhood observation
is simply because we cannot resolve 
the bulge and the disk into separate components.

\section{Conclusions}
\label{conclusion}
We have studied predictions of the $\Lam$CDM universe 
with regard to the star formation history and the stellar 
metallicity distribution in galaxies. 
Our purpose has been to clarify what observations
would test the validity of the model of galaxy formation 
based on $\Lambda$CDM. 
We have first shown that the global SFR averaged over 
all galaxies declines with a characteristic time-scale of
$\tau\approx 6\Gyr$. 
About a quarter of stars formed by $z\approx 3.5$ and
another quarter by $z\approx 2$. This star formation history 
is consistent with the empirical Madau plot if modest 
dust obscuration is assumed.

Our calculation indicates that star formation in $L^*$ galaxies 
continues {\it intermittently}
to the present epoch as they accrete gas and merge with smaller systems
with a decline rate corresponding to an e-folding time of $\tau\approx 6\Gyr$. 
Star formation in less massive galaxies ceases earlier
with an e-folding time of $\tau \ltsim 1\Gyr$.
In particular, dwarf galaxies cease their star formation by $z=2$. 
It is not excluded that minor star formation activity at low-redshift
is missed in our calculation due to yet insufficient resolution, 
but it is unlikely that the global trend would be modified.
We consider that this is a feature generic to the CDM
structure formation scenario, and observational tests would 
provide a vital test.

Stars formed in the CDM galaxy formation scenario do not follow a
unique age-metallicity relation, but show a considerable spread of
metallicity for a given age of stars. Metallicity of very young stars 
spreads from $0.1-1.0\Zsun$. 
The spread gradually increases towards high redshift. 
It takes $0.01-1.0\Zsun$ at $z=2$.
At $z=3$, the distribution spreads over from $10^{-6}\Zsun$ 
to super-metal rich ($3\Zsun$). 
Young stars are necessarily metal rich, 
but old stars are not necessarily metal poor. 
The variation of the mean metallicity with cosmic time is only gradual. 
The average metallicity for galaxies at $z\approx 1$
differs little from that of local galaxies.

On the other hand, mean metallicity varies as a function of 
the mass of galaxies. Less massive galaxies are metal poor. 
Another interesting prediction is that mean metallicity of 
dwarf galaxies today ranges widely ($10^{-4}-1\Zsun$), 
which reflects the large scatter of metallicity of
stars formed at high-redshift.

The G-dwarf problem in the solar neighborhood offers an 
interesting insight for formation of galactic structure. 
If we apply the prediction of the metallicity distribution 
to solar neighborhood, we encounter the G-dwarf problem: 
the prediction of the metallicity distribution 
does not differ much from that of the closed-box model.
In our case, however, the G-dwarf problem would be solved if 
bulges are early entities and disks are later additions.

We have briefly discussed observational tests for each item of the
predictions. In many aspects the predictions are supported by
observational evidence, but for some aspects the predictions
do not seem to agree with the current interpretation of the 
observations.
The currently available observations, however, are sometimes 
controversial among authors or do not offer unique interpretations 
free from various working assumptions. 
We consider that decisive tests will be a task for the
future.

\acknowledgments
We thank Simon White for suggesting that we check the stability of 
the galaxies in our simulation.
We are grateful to the anonymous referee and Michael Strauss for 
useful comments on the draft. 
K.N. is supported in part by the Physics Department of Princeton University.
M.F. is supported in part by the Raymond and Beverly Sackler Fellowship 
in Princeton and Grant-in-Aid of the Ministry of Education of Japan.
R.C. and J.P.O. are partly supported by grants AST~98-03137 and ASC~97-40300.

%----------------------------------------------------------------------


\begin{thebibliography}{}
\bibitem[Alonso-Herrero~\etal(1996)]{Alonso-Herrero96} Alonso-Herrero, A., Aragon-Salamanca, A., Zamorano, J., \& Rego, M. 1996, \mnras, 278, 417 
\bibitem[Arnett(1996)]{Arnett96} Arnett, D., 1996, {\it Supernovae and Nucleosynthesis}, Princeton University Press, pp.496
%\bibitem[Bahcall, Ostriker, and Steinhardt(1999)]{Bahcall99} Bahcall, N. A., Ostriker, J. P., and Steinhardt, P. J. 1999, Science, 284, 1481
\bibitem[Balbi~\etal(2000)]{Balbi00} Balbi, A., ~\etal~2000, \apj, 545, L1 %MAXIMA-I 
\bibitem[Baugh~\etal(1998)]{Baugh98} Baugh, C. M., Cole, S., Frenk. C. S.,~\& Lacey, C.~G., 1998 ApJ, {498}, 504
%\bibitem[Binney and Tremaine(1987)]{Binney87} Binney, J. and Tremaine, S. 1987, {\it Galactic Dynamics}, Princeton University Press, Princeton
%\bibitem[Binney and Merrifield(1998)]{Binney98} Binney, J. and Merrifield, M. 1998, {\it Galactic Astronomy}, Princeton University Press, Princeton
%\bibitem[Blanton~\etal(1999)]{Blanton99} Blanton, M., Cen, R., Ostriker, J. P., and Strauss, M. A.  1999, \apj, 522, 590 
%\bibitem[Blanton \etal(2000)]{Blanton00} Blanton, M., Cen, R., Ostriker, J. P., Strauss, M. A., and Tegmark, M.  2000, \apj, 531, 1
\bibitem[Blumenthal~\etal(1984)]{Blumenthal84} Blumenthal, G. R., Faber, S. M., Primack, J. R., \& Rees, M. J. 1984, \nat, 311, 517  
\bibitem[Bode \etal(2000)]{Bode00} Bode, P., Ostriker, J. P., and Turok, N., preprint (astro-ph/0010389)
%\bibitem[Bond(1970)]{Bond70} Bond, H. E. 1970, \apjs, 22, 117
\bibitem[Bruzual \& Charlot(1993)]{Bruzual93} Bruzual, A.~G.~\& Charlot, S., 1993, ApJ, {405}, 538
\bibitem[Bryan \& Norman(1995)]{Bryan95} Bryan, G. L. \& Norman, M. L. 1995, AAS, 187.9504
%\bibitem[Carollo and Lilly(2000)]{Carollo00} Carollo, C. M. and Lilly, S. J. 2000, in press (preprint astro-ph/0011353)
\bibitem[Cen and Ostriker(1992a)]{CO92a} Cen, R. and Ostriker, J. P.  1992a, \apj, 393, 22
%\bibitem[Cen and Ostriker(1992b)]{CO92b} Cen, R. and Ostriker, J. P.  1992b, \apjl, 399, L113
\bibitem[Cen and Ostriker(1993)]{CO93} Cen, R. and Ostriker, J. P.  1993, \apj, 417, 404
%\bibitem[Cen(1998)]{Cen98} Cen, R.  1998, \apj, 509, 16
%\bibitem[Cen and Ostriker(1999a)]{CO99a} Cen, R. and Ostriker, J. P.  1999a, \apj, 514, 1 %where are the baryons
%\bibitem[Cen and Ostriker(1999b)]{CO99b} Cen, R. and Ostriker, J. P.  1999b, \apj, 517, 31 %accuracy
\bibitem[Cen and Ostriker(1999c)]{CO99c} Cen, R. and Ostriker, J. P.  1999c, \apjl, 519, L109 %cosmic chemical evolution
\bibitem[Cen and Ostriker(2000)]{CO00} Cen, R. and Ostriker, J. P.  2000, \apj, 538, 83 %physical bias
%\bibitem[Clegg and Bell(1973)]{Clegg73} Clegg, R. E. S. and Bell, R. A. M. 1973, \mnras, 163, 13
\bibitem[Cole~\etal(2000)]{Cole00} Cole, S., Lacey, C. G., Baugh, C. M., \& Frenk, C. S. 2000, \mnras, 319, 168
%\bibitem[Connolly~\etal(1997)]{Connolly97} Connolly,~A.~J., Szalay,~A.~S., Dickinson,~M.~E., SubbaRao,~M.~U.,~\&~Brunner,~R.~J., 1997, \apj, 486, L11 
%\bibitem[Dave~\etal(2000)]{Dave00} Dave, R., Hernquist, L., Katz, N., \& Weinberg, D. H. 2000, Clustering at High Redshift, ASP Conference Series, Vol. 200. Edited by A. Mazure, O. Le Fvre, and V. Le Brun, p.402
\bibitem[Dave, Dubinski, \& Hernquist(1997)]{Dave97} Dave, R., Dubinski, J., \& Hernquist, L. 1997, New Astronomy, 2, 277
\bibitem[Davis~\etal(1985)]{Davis85} Davis, M., Efstathiou, G., Frenk, C. S., \& White, S. D. M. 1985, \apj, 292, 371
\bibitem[Dekel \& Silk(1986)]{Dekel86} Dekel, A. and Silk, J., \apj, 1986, 303, 39
\bibitem[Edelson and Malkan(1986)]{Edelson86} Edelson, R. A. and Malkan, M. A. 1986, \apj, 308, 59
\bibitem[Edvardsson~\etal(1993)]{Edvardsson93} Edvardsson, B., Andersen, J., Gustafsson, B., Lambert, D. L., Nissen, P. E., \& Tomkin, J. 1993, \aap, 273, 101
\bibitem[Efstathiou, Sutherland, \& Maddox(1990)]{Efstathiou90} Efstathiou, G., Sutherland, W. J., \& Maddox, S. J. 1990, \nat, 348, 705
\bibitem[Efstathiou(1992)]{Efstathiou92} Efstathiou, G., 1992, \mnras, 256, 43
\bibitem[Eisenstein and Hut(1998)]{Eisenstein98} Eisenstein, D. J.~and~Hut, P.  1998, \apj, 498, 137
\bibitem[Fan~\etal~(2000)]{Fan00} Fan, X., \etal~2000, AJ, 120, 1167
%\bibitem[Fardal~\etal(2000)]{Fardal00} Fardal, M., ~\etal, 2000, preprint (astro-ph/0007205)
\bibitem[Frogel \& Whitford(1987)]{Frogel87} Frogel, J. A. and Whitford, A. E. 1987, \apj, 320, 199 
\bibitem[Fukugita, Hogan, and Peebles(1998)]{Fukugita98} Fukugita, M., Hogan, C. J., Peebles, P. J. E., 1998, ApJ, {503}, 518 
\bibitem[Fukugita, Hogan, \& Peebles(1996)]{Fukugita96} Fukugita, M., Hogan, C. J., Peebles, P. J. E., 1996, Nature, {381}, 489 
\bibitem[Gallagher, Hunter, \& Tutukov(1984)]{Gallagher84} Gallagher, J. S., Hunter, D. A., Tutukov, A. V. 1984, \apj, 284, 544 
%\bibitem[Gallego \etal(1996)]{Gallego96} Gallego, J., Zamorano, J., Arag\`{o}n-Salamanca, \& Rego, M., 1995, \apj, 455, L1; erratum, 459, L43 (1996)
%\bibitem[Garnavich \etal (1998)]{Garnavich98} Garnavich, P. M., \etal  1998, \apj, 509, 74
\bibitem[Gnedin(2000b)]{Gnedin00b} Gnedin, N. Y. 2000b, \apj, 542, 535 %effect of photoionization
\bibitem[Gnedin(2000a)]{Gnedin00a} Gnedin, N. Y. 2000a, \apj, 535, L75 %dwarf gals
\bibitem[Gnedin(1998)]{Gnedin98} Gnedin, N. Y. 1998, \apj, 294, 407 %metal enrichment
\bibitem[Grebel(1997)]{Grebel97} Grebel, E. K. 1997, Reviews in Modern Astronomy, 10, 29
\bibitem[Hu,~\etal(2000)]{Hu00} Hu, W., Fukugita, M., Zaldarriaga, M., \& Tegmark, M. 2001, \apj, 549, 669 
\bibitem[Kauffmann~\etal(1999a)]{Kauffmann99a} Kauffmann, G., Colberg, J. M., Diaferio, A., \& White, S. D. M., 1999a, MNRAS, 303, 188 
%\bibitem[Kauffmann~\etal(1999b)]{Kauffmann99b} Kauffmann, G., Colberg, J. M., Diaferio, A., \& White, S. D. M., 1999b, MNRAS, 307, 529 
\bibitem[Kauffmann(1996)]{Kauffmann96} Kauffmann, G. 1996, \mnras, 281, 475
\bibitem[Katz, Weinberg, \& Hernquist(1996)]{Katz96} Katz, N., Weinberg, D. H., \& Hernquist, L. 1996, \apjs, 105, 19
\bibitem[Kobulnicky and Zaritsky(1999)]{Kobulnicky99} Kobulnicky, H. A. and Zaritzky, D. 1999, \apj, 1999, 511, 118
\bibitem[Kravtsov, Klypin, \& Khokhlov(1997)]{Kravtsov97} Kravtsov, A. V., Klypin, A. A., \& Khokhlov, A. M. 1997, \apjs, 111, 73
\bibitem[Lange~\etal(2000)]{Lange00} Lange, A. E., ~\etal~2000, preprint (astro-ph/0005004) %BOOMERANG
\bibitem[Lanzetta~\etal(1999)]{Lanzetta99} Lanzetta, K. M., \etal ~1999, the proceedings of ``The Hy-Redshift Universe: Galaxy Formation and Evolution at High Redshift'', ASP Conference Series, Vol. 193, Ed. A. J. Bunker and W. J. M. van Breugel, p.544
%\bibitem[Lilly~\etal(1996)]{Lilly96} Lilly, S. J., Le F\`{e}vre, O., Hammer, F., \& Crampton, D., 1996, \apj, 460, L1 
%\bibitem[Lowenthal~\etal(1997)]{Lowenthal97} Lowenthal, J. D., \etal 1997, \apj, 481, 673
%\bibitem[Madau~\etal(1996)]{Madau96} Madau, P., \etal ~1996, \mnras, 283, 1388  
%\bibitem[Madau(1997)]{Madau97} Madau,~P., 1997, in AIP Conf. Proc. 393, Star Formation Near and Far, ed. S.~S.Holt~\& G.~L.~Mundy (New York:AIP), 481 
\bibitem[Majewsky(1993)]{Majewsky93} Majewsky, S. R. 1993, ARA\&A, 31, 575
\bibitem[Mateo(1998)]{Mateo98} Mateo, M. 1998, ARA\&A, 36, 435
\bibitem[McWilliam \& Rich(1994)]{McWilliam94} McWilliam, A. \& Rich, R. M. 1994, \apjs, 91, 749
\bibitem[McWilliam(1997)]{McWilliam97} McWilliam, A. ARA\&A, 1997, 35, 503
\bibitem[Nagamine, Fukugita, Cen, and Ostriker(2001)]{Nagamine01} Nagamine, K., Fukugita, M., Cen, R., \& Ostriker, J. P. 2001, preprint (astro-ph/0102180; Paper II) 
%\bibitem[Nagamine, Ostriker and Cen(2000)]{Nagamine00b} Nagamine, K., Ostriker, J. P., and Cen, R. 2000, \apj, in press (astro-ph/0010253) %Mach number 
\bibitem[Nagamine~\etal(2000)Nagamine, Cen, and Ostriker]{Nagamine00a} Nagamine, K., Cen, R., and Ostriker, J. P.  2000, \apj, 541, 25  %luminosity density
%\bibitem[Nagamine, Cen, and Ostriker(1999)]{Nagamine99} Nagamine, K., Cen, R., \& Ostriker, J. P. 1999, the proceedings of the 4th RESCEU International Symposium: "The Birth and Evolution of the Universe", in press (astro-ph/9912023)
\bibitem[Navarro \& Steinmetz(1997)]{Navarro97} Navarro, J. F. \& Steinmetz, M. 1997, \apj, 478, 13
%\bibitem[Norman(2000)]{Norman00} Norman, M. L. 2000, private communication
\bibitem[Ostriker and Steinhardt(1995)]{Ostriker95} Ostriker, J. P. and Steinhardt, P. J.  1995, \nat, 377, 600
\bibitem[Ostriker and Thuan(1975)]{Ostriker75} Ostriker, J. P. and Thuan, T. X. 1975, \apj, 202, 353
\bibitem[Pagel \& Patchett(1975)]{Pagel75} Pagel, B. E. J. and Patchett, B. E. 1975, \mnras, 172, 13 
%\bibitem[Pagel(2001)]{Pagel01} Pagel, B. E. J. 2001, to appear in `Cosmic Evolution' Conference at IAP, Paris, preprint (astro-ph/0101376)
%\bibitem[Pascarelle, Lanzetta, \& Fernandez-Soto(1998)]{Pascarelle98} Pascarelle, S. M., Lanzetta, K. M., \& Fernandez-Soto, A., 1998, \apj, 508, L1
\bibitem[Pearce \etal(1999)]{Pearce99} Pearce, F. R., \etal ~1999, \apjl, 521, L99
%\bibitem[Peebles(2000)]{Peebles00} Peebles, P. J. E. 2000, in IAU Conference 204, the Extragalactic Infrared Background and its Cosmological Implications, Manchester, eds M. Harwit and M. G. Hauser (astro-ph/0010617)
%\bibitem[Peebles(1980)]{Peebles80} Peebles, P. J. E. 1993, {\it Large-scale Structure of the Universe} (Princeton: Princeton University Press)
%\bibitem[Peebles(1993)]{Peebles93} Peebles, P. J. E. 1993, {\it Principles of Physical Cosmology} (Princeton: Princeton University Press)
%\bibitem[Perlmutter \etal(1998)]{Perlmutter98} Perlmutter, S., et al.  1998, \nat, 391, 51
\bibitem[Pettini~\etal(2000)]{Pettini00} Pettini, M., Steidel, C. C., Adelberger, K. L., Dickinson, M., \& Giavalisco, M. 2000, \apj, 528, 96
\bibitem[Phillips, Ostriker, \& Cen(2000)]{Phillips00} Phillips, L. A., Ostriker, J. P., \& Cen, R. 2000, preprint (astro-ph/0011348)
\bibitem[Quinn, Katz, \& Efstathiou(1996)]{Quinn96} Quinn, T., Katz, N., \& Efstathiou, G. 1996, \mnras, 278, L49
\bibitem[Rees(1986)]{Rees86} Rees, M. J., \mnras, 1986, 218, 25 
%\bibitem[Reichardt, Jimenez, and Heavens(2001)]{Reichardt01} Reichardt, C., Jimenez, R., and Heavens, A. F. 2001, preprint (astro-ph/0101074)
%\bibitem[Riess \etal(1998)]{Riess98} Riess, A.~G., et al.  1998, \aj, 116, 1009
\bibitem[Rich(1988)]{Rich88} Rich, R. M. \aj, 95, 828
\bibitem[Rocha-Pinto \etal(2000a)]{Rocha-Pinto00a} Rocha-Pinto, H. J., Maciel, W. J., Scalo, J., \& Flynn, C. 2000a, \aap, 358, 850
\bibitem[Rocha-Pinto \etal(2000b)]{Rocha-Pinto00b} Rocha-Pinto, H. J., Scalo, J., Maciel, W. J., \& Flynn, C. 2000b, \aap, 358, 869
%\bibitem[Ronen, Arag\'{o}n-Salamanca, \& Lahav(1999)]{Ronen99} Ronen, S., Arag\'{o}n-Salamanca, \& Lahav, O. 1999, \mnras, 303, 284 
\bibitem[Ryu~\etal(1993)]{Ryu93} Ryu, D., Ostriker, J. P., Kang, H., and Cen, R.  1993, \apj, 414, 1
\bibitem[Sandage(1986)]{Sandage86} Sandage, A. 1986, \aap, 161, 89
%\bibitem[Sawicki, Lin, \& Yee(1997)]{Sawicki97} Sawicki, M. J., Lin, H., \& Yee, H. K. C., \aj, 113, 1
\bibitem[Scalo(1986)]{Scalo86} Scalo, J. N., 1986, Fundam. Cosmic. Phys., 11, 1
\bibitem[Schmidt(1963)]{Schmidt63} Schmidt, M. 1963, \apj, 137, 758
%\bibitem[Searle, Sargent, \& Bagnuolo(1973)]{Searle73} Searle, L., Sargent, W. L. W., and Bagnuolo, W. G. 1973, \apj, 179, 427
\bibitem[Sommer-Larsen(1991)]{Sommer-Larsen91} Sommer-Larsen, J. 1991 \mnras, 249, 368
\bibitem[Somerville, Primack, \& Faber(2001)]{SP01} Somerville, R. S., Primack, J. R., \& Faber, S. M. 2001, MNRAS, 320, 504
\bibitem[Somerville and Primack(1999)]{SP99} Somerville, R. S., and Primack, J. R., 1999, MNRAS, 310, 1087
\bibitem[Steidel~\etal(1999)]{Steidel99} Steidel, C. C., Adelberger, K. L., Giavalisco, M., Dickinson, M., Pettini, M., 1999, \apj, 519, 1 
%\bibitem[Treyer~\etal(1998)]{Treyer98} Treyer, M. A., Ellis, R. S., Millard, B., Donas, J, \& Bridges, T. J., 1998, \mnras, 300, 303
%\bibitem[Tresse \& Maddox(1998)]{Tresse98} Tresse, L. \& Maddox, S. J., 1998, \apj, 495, 691
\bibitem[Turner and White(1997)]{Turner97} Turner, M. S. and White, M. 1997, Phys. Rev. D, 56, 4439
\bibitem[van den Bergh(1962)]{Bergh62} van den Bergh, S. 1962, \aj, 67, 486
\bibitem[van den Bergh(1998)]{Bergh98} van den Bergh, S. 1998, A\&A review, 9, 273
%\bibitem[Wang~\etal(2000)]{Wang00} Wang, L., Caldwell, R. R., Ostriker, J. P., and Steinhardt, P. J.  2000, \apj, 530, 17
\bibitem[Weinberg, Hernquist, \& Katz(1997)]{Weinberg97} Weinberg, D. H., Hernquist, L., \& Katz, N. 1997, \apj, 477, 8
\bibitem[White \& Frenk(1991)]{White91} White, S. D. M. \& Frenk, C. S. 1991, \apj, 379, 52
%\bibitem[Williams~\etal(1996)]{Williams96} Williams R. E., et al., 1996, \aj, 112, 1335

%   \aap   A&A
%   \aj    AJ
%   \pasp  PASP
%   \mnras MNRAS
%   \apjs  ApJS
\end{thebibliography}
\end{document}